\documentclass[prl,twocolumn,amsmath,amssymb,superscriptaddress,longbibliography]{revtex4-1}
\usepackage{epsfig, graphicx,graphics,amsmath,amssymb,float}
\usepackage[T1]{fontenc}
\usepackage[latin9]{inputenc}
\usepackage{amsmath}
\usepackage{amssymb}
\usepackage{xcolor}
\usepackage{amscd}
\usepackage{bm}
\usepackage{bbold}
\usepackage{psfrag}

\usepackage{bbm}
\usepackage{algorithm2e}
\usepackage{graphicx}

\usepackage[normalem]{ulem} 

\usepackage[caption=false]{subfig}
\usepackage{subfig}
\usepackage{color}
\usepackage{xurl}            
\usepackage[bookmarks=true,colorlinks,linkcolor=blue,urlcolor=blue,citecolor=blue]{hyperref}

\begin{document}

\title{
  Pseudogap-induced change in the nature of the Lifshitz transition in the two-dimensional Hubbard model}

\author{Maria C. O. Aguiar}%
\affiliation{Departamento de F\'isica, Universidade Federal de Minas Gerais,
  C.P.~702, 30123-970, Belo Horizonte, MG, Brazil}%
\affiliation{Universit\'e Paris-Saclay, CNRS, Laboratoire de Physique des
  Solides, 91405, Orsay, France}%
\author{Helena Bragan\c{c}a}
\affiliation{Instituto de F\'{i}sica and International Center for Physics, Universidade de Bras\'{i}lia, Bras\'{i}lia 70919-970, DF, Brazil}%
\author{Indranil Paul}
\affiliation{Universit\'e de Paris, Laboratoire Mat\'eriaux et Ph\'enom\`enes
Quantiques (UMR 7162 CNRS), Bat. Condorcet, 75205 Paris Cedex 13, France}
\author{Marcello Civelli}%
\affiliation{Universit\'e Paris-Saclay, CNRS, Laboratoire de Physique des
  Solides, 91405, Orsay, France}%

\date{\today}

\begin{abstract}

We study the behavior of the density of states and the B$_{1g}$ nematic susceptibility extracted from Raman response data across the doping-driven Lifshitz transition comparing the weak and strong interaction cases. Our results were obtained using cluster dynamical mean field theory for the two-dimensional Hubbard model. In the weakly correlated Fermi liquid regime, both quantities are approximately symmetric around the Lifshitz transition doping $p_{LT}$. In the strongly correlated regime, the low-doping pseudogap leads to an asymmetric, discontinuous evolution when the Fermi surface changes from hole-like to electron-like at $p_{LT}$. The Lifshitz transition thus changes character because it is tied to the pseudogap-Fermi-liquid transition. These results are consistent with available observations and should foster further experimental investigations.
\end{abstract}

\maketitle

\section{Introduction} \label{intro}

Despite all the efforts of recent years, the origin of the
superconductivity~\cite{bcs,eliashberg,stewart_2017} observed
in copper oxides (cuprates) remains an open problem. The normal
metallic state above the superconducting critical
temperature T$_{\mathrm{C}}$ -- the pseudogap phase~\cite{alloul_1989,warren_1989}, from which superconductivity emerges -- is also currently not well understood. This phase is
characterized by a loss of spectral weight at some regions of
the momentum space, as seen in spectroscopic, thermodynamic, and
transport measurements~\cite{timusk_1999,damascelli_2003,piriou_2011}.
Clarifying the role of the pseudogap in the high-temperature 
superconducting mechanism remains one of the central open questions 
in the field~\cite{Norman2005Pseudogap}.

In the pioneering experiment of Ref.~\cite{piriou_2011}, 
in particular, scanning tunneling spectroscopy (STS) was used on  
Bi$_2$Sr$_2$CuO$_{6+\delta}$ above T$_{\mathrm{C}}$. Due to intrinsic surface 
inhomogeneities, the local doping changes as the STS tip moves, 
allowing one to probe regions with different doping levels. As the 
tip scans across the surface, the spectra evolve from those displaying 
a pseudogap spectral weight suppression to spectra without 
the pseudogap, displaying instead a prominent peak at the Fermi level.
This observation suggests the proximity to a Van Hove singularity (VHS), 
i.e. a Lifshitz transition.

More recently,
experiments on cuprates have pointed out that the doping $p^*$
at which the pseudogap ends coincides with the critical doping
$p_{LT}$ of a Lifshitz transition from a hole-like to an
electron-like Fermi surface
(FS)~\cite{benhabib_2015,loret_2017,doiron_2017}.
In a previous work by some of us~\cite{helena_prl}, we have
theoretically addressed this issue and found this concurrence
in the regime of strong interactions. To be more specific, we have
explored the phase diagram of two-dimensional systems described
by the doped Hubbard model and treated within the cellular
dynamical mean field theory (CDMFT)~\cite{Maier05,BGK06}. 
At weak interactions, we
find that the system is always in correlated metallic ground state, a Fermi liquid (FL),
which goes through the Lifshitz transition cited above as doping is increased.
At strong interactions, region that is relevant to the cuprates,
the pseudogap phase becomes the ground state at small dopings.
By increasing doping, we observe a first-order transition from
the pseudogap to the conventional FL metal, which is tied to the
change of the FS from hole- to electron-like, that is, at
large interactions we obtain that $p^* \approx p_{LT}$, in
accordance with the experiments~\cite{benhabib_2015,loret_2017,doiron_2017}.

The VHS~\cite{doiron_2017} and the coincidence between $p^*$ and  $p_{LT}$ have been observed experimentally, in particular, through Raman spectroscopy~\cite{benhabib_2015,loret_2017}.
As this probes the electron dynamics in different regions of the Brillouin zone (BZ), the Raman response has shed light on the impact of correlations and competing orders on high-temperature superconductivity~\cite{rmp_raman}.
Symmetry-resolved electronic Raman scattering has been used, for instance, to investigate the effects of nematic fluctuations on the cuprate phase diagram -- these fluctuations are related to a breaking of the C$_4$ rotational symmetry of the Cu square lattice~\cite{natphys_gallais}.

Here, motivated by the possibility of a comparison between our theoretical investigation and these measurements, we continue to investigate the system considered in Ref.~\cite{helena_prl} by analyzing the behavior
of the density of states (DOS) and the nematic susceptibility extracted from Raman response data across the interaction versus doping phase diagram. 
We show that the doping dependence of these two quantities are
qualitatively different in the strongly correlated regime
as compared to the weakly one. While in the latter the DOS at
the Fermi level and the nematic susceptibility are approximately
symmetrical around $p_{LT}$, for the former we observe a 
discontinuous and asymmetrical behavior. We ascribe this
change in the nature of the Lifshitz transition to the fact that
at strong interactions it is tied to the transition from the
pseudogap to the FL phase. Within the pseudogap
phase, we also analyze how the FS evolves from a large hole-like one
close to $p_{LT}$ into a small elliptical pocket at small doping.
 
The paper is organized as follows. In the next section we
define the model we consider and give details on the
methodology used to solve it. In Sec.~\ref{results}, we
present our numerical results. Firstly, we focus on the DOS 
at the Fermi level and the nematic susceptibility as a function of
doping, discussing the qualitatively different behavior
observed at weak and strong interactions. Secondly, we discuss
the evolution of the FS within the pseudogap phase. Finally, 
Sec.~\ref{conclusions} is devoted to our conclusions.

\section{Model and Methodology}

We consider the two-dimensional (2D) one-band Hubbard model on
a $L$-site square lattice, which is given by the Hamiltonian
\begin{eqnarray} \label {Hm}
  \mathcal{H}=\, - \sum_{ \mathbf{k}  \sigma} \, \xi_{\mathbf{k}} \, c^\dagger_{\mathbf{k}\sigma}
  c_{\mathbf{k} \sigma}  + U \sum_{i } n_{i \uparrow}n_{i \downarrow}, \label{model}
\end{eqnarray}
where 
$c_{\mathbf{k} \sigma}= (1/\sqrt{L}) \sum_{i} \exp(-\textit{i}
\mathbf{k} \cdot \mathbf{r}_i) \, c_{i\sigma}$ destroys an
electron with spin $\sigma$ and momentum $\mathbf{k}$,   
$n_{i \sigma}=c^\dagger_{i \sigma}c_{i \sigma}$ is the density
operator on site $i$ of the lattice, and $U$ is the onsite
Coulomb repulsion. The electronic dispersion is given by
\begin{equation}
  \xi_{\mathbf{k}}= -2t (\cos k_x + \cos k_y)- 4t^{\prime} \cos k_x \cos k_y- \mu,
  \label{dispersion}
\end{equation}
where $t$ and $t^{\prime}$ are nearest- and
next-nearest-neighbor-site hopping integrals, respectively,
and $\mu$ is the chemical potential, through which
the doping level $p= 1- (1/L) \sum_{i \sigma} \langle
n_{i\sigma} \rangle$ is controlled. Throughout the paper
we set $t=1$ as the unit of energy and consider $t'=-0.1$.

To solve the model defined in Eq.~(\ref{model}), we use
CDMFT, which well captures the phases appearing in the
cuprate phase diagram, including the Mott insulator, the
antiferromagnetic phase, the pseudogap state, and the
superconducting
phase~\cite{Maier05,BGK06,AMT06,SSK08,ferrero09,AMT10, Gull10,Gull13,Gull15}.
Here, we focus on the normal, paramagnetic solution
at zero temperature and in the presence of doping $p$.
We are particularly interested in the pseudogap regime. Therefore, we explicitly consider the normal, non-superconducting solution of the Hubbard model. This phase is representative of the large pseudogap region observed above the superconducting critical temperature in the cuprate phase diagram, although our calculations are performed at $T=0$.
We stress that the theoretical doping values $p$ are smaller
than the typical experimental dopings observed in cuprates.

Within the CDMFT implementation we consider, the model
defined in Eq. (\ref{model}) is mapped onto a $2 \times 2$
cluster embedded in a $8$-site bath 
(an Anderson cluster-impurity model), which is solved by
exact diagonalization (ED). The CDMFT + ED calculation
is performed on the imaginary axis, with fictitious temperature equal to $T/t= 0.01$.
Previously~\cite{helena_prl}
we have used a reduced bath parametrization, while in the
present work we consider a more general description -- the
relaxed bath parametrization; both schemes are
described in the supplemental material of Ref.~\cite{helena_prl}. 
The numerical CDMFT calculation provides
frequency dependent quantities, such as the Green's function
$G(\mathbf{k},i\omega)$,
in symmetric points $\mathbf{k}=(0, \pi), \ (\pi, 0), \ (\pi,\pi)$
and $(0,0)$ of the first quadrant of the BZ.
To obtain the physical lattice quantities in momentum space we perform
a periodization based on the cumulant~\cite{PM,PM2}; see
Appendix~\ref{cumulant} for details on this procedure. 

In the next section, we present our numerical results for the
local DOS,
$\rho_{loc}(i\omega)=(-1/\pi) \sum_{\mathbf{k}} \mbox{Im} G(\mathbf{k},i\omega)$,
close to the Fermi level, as well for the nematic susceptibility extracted from the Raman response,
calculated as we detail below.

\subsection{Nematic susceptibility}

From the Matsubara Green's functions $G(\mathbf{k},i\nu_n)$ obtained
after the periodization mentioned above, we can compute the
Raman response on the imaginary axis within first order perturbation theory as follows

\begin{equation}
\chi(i\omega)=- \frac{2}{\beta} \sum_{\mathbf{k} \nu_n}
\gamma^2(\mathbf{k}) G(\mathbf{k},i\nu_n) G(\mathbf{k},i\nu_n+i\omega),
\label{chiiom}
\end{equation}
where the factor $2$ accounts for the spin degeneracy and $\gamma(\mathbf{k})$ is the Raman scattering vertex amplitude,
which depends on the symmetry. Here we investigate the $B_{1g}$ scattering, that corresponds to cross-photon polarizations at 45 degrees of
the Cu-O-Cu direction and probes the doping effect on the oscillator strength in the $(\pi, 0)$ regions of the BZ~\cite{rmp_raman}. In such case, the scattering amplitude 
is given by
\begin{equation}
  \gamma_{B1g}(\mathbf{k})=2t (\cos k_x - \cos k_y ).
  \label{ramanB1g}
\end{equation}
See Appendix~\ref{app_raman} for more details on this Raman
factor.

We can express the Raman response on the real axis by
considering the analytical continuation of the expression
in Eq.~(\ref{chiiom}). The two quantities -- the one on the
imaginary axis and that on the real axis -- are related
through the Hilbert transform:
\begin{equation}
\chi(i\omega)= -\frac{1}{\pi} \int_{-\infty}^{\infty} d\nu \frac{\chi''(\nu)}{i\omega-\nu},
\end{equation}
where $''$ denotes the imaginary part.
We are particularly interested in the $i \omega \rightarrow 0$
limit, where we can write
\begin{equation}
  \chi(i\omega \rightarrow 0)= \frac{1}{\pi} \int_{-\infty}^{\infty} d\nu \frac{\chi''(\nu)}{\nu}.
  \label{hilbertw0}
\end{equation}
$\chi''(\nu)$ can be directly compared to the experimental measurements.
An integral similar to the one in Eq.~(\ref{hilbertw0}),
with limits of integration from $0$ to a cutoff $\Lambda$, was
calculated from the experimental results in Ref.~\cite{natphys_gallais}
and associated to a nematic susceptibility originating from a broken
rotational symmetry in the CuO$_2$ plane of cuprates.
Numerically, it is more precise to obtain $\chi(i\omega \rightarrow 0)$ 
on the imaginary axis than calculate the integral on the right side of
Eq.~(\ref{hilbertw0}). The reason for it is that the CDMFT + ED calculation
is performed on the imaginary axis and to obtain $\chi''(\nu)$ one needs to
perform an analytic continuation $i\nu_n\to \nu+ i \eta$ using a small but finite
broadening $\eta$. In the following section, we will analyze the results we
have obtained for the nematic susceptibility
$\chi(i\omega \rightarrow 0)$ as a function of doping in the weakly
and in the strongly correlated regimes.

\section{Numerical Results} \label{results}

\subsection{Lifshitz transition nature for weak and strong interactions}
\label{DOSRaman} 

In this section we present our results for the DOS at the
Fermi level as well as for the nematic susceptibility extracted from the
Raman response as a function of doping
obtained for $U = 3.5$ and $U=8.0$. We will see that the results at
weak correlation, corresponding to the system in the FL phase, are
qualitatively different than the ones at strong interaction, in which
the system goes from the pseudogap phase at small doping to the FL
phase at large doping.

\subsubsection{Density of states results}

\begin{figure}[t]
\begin{center}
  \includegraphics[scale=0.33]
  {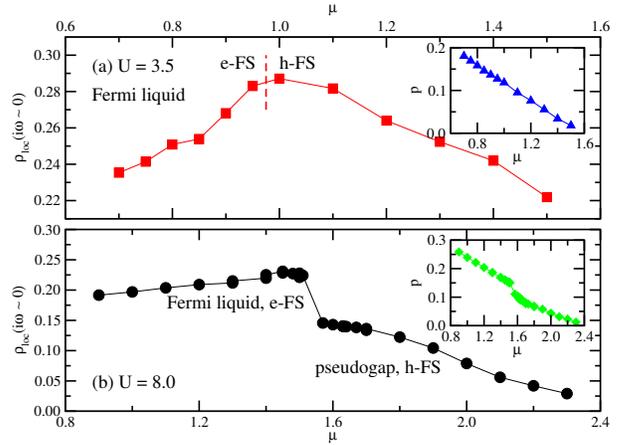}
\end{center}
\caption{
  Local DOS at the Fermi level as a function of the chemical potential $\mu$ for (a) $U=3.5$
  and (b) $U=8.0$. The insets show the doping $p$ as a function of $\mu$ corresponding
  to data in the respective main panel. The dashed red line in panel (a) marks the Lifshitz
  transition. Note that the DOS for $U=8.0$ is always smaller and varies in a
  broader range than it is the case for $U=3.5$.
}
\label{DOSvsmu}
\end{figure}

Our theoretical results for the local DOS at the Fermi level $\rho_{loc}(i\omega \rightarrow0)$
as a function of the chemical potential are presented in
Fig.~\ref{DOSvsmu} in the weak
(a) $U=3.5$ and strongly correlated (b)~$U=8.0$ regimes. 
The insets show the dependence
of the doping $p$ with the chemical potential $\mu$ for each case. 
The full local DOS on a broader
real frequency region for some $\mu$ values is presented in the Appendix~\ref{DOScomparison}.
The first striking observation about Fig.~\ref{DOSvsmu} is that 
while $\rho_{loc}(i\omega \rightarrow0)$
is approximately symmetrical around its highest value for $U=3.5$, 
no such a symmetry and a discontinuity are observed for $U=8.0$. 
A discontinuity is also seen in the $p$ vs. $\mu$ curve
for $U=8.0$, which is not present for $U=3.5$. Within our numerical
treatment, this is a strong indication of (or a proximity to) a first
order transition.

\begin{figure}[t]
\begin{center}
  \includegraphics[scale=0.45]
                 {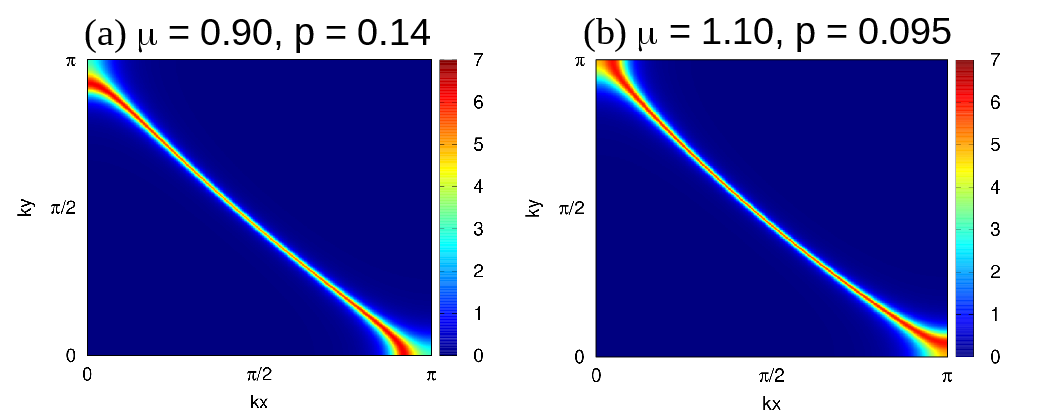}
\end{center}
\caption{Spectral function $A(\mathbf{k}, iw \rightarrow 0$) for $U=3.5$ and values of the chemical potential/doping close to the Lifshitz transition. The system is in the FL phase in both cases; in panel (a) it has an e-FS, while in (b) it has a h-FS. 
}
\label{AkwU35}
\end{figure}

We have analyzed the spectral function $A(\mathbf{k},i\omega)=
-(1/\pi) \mbox{Im} G(\mathbf{k},i\omega)$ and the corresponding
self-energy $\Sigma(\mathbf{k},\omega)$ at $\mathbf{k}=(0,\pi)$ 
(as we did previously in Ref.~\cite{helena_prl}) and have concluded that for $U=3.5$ the system ground state
presents a FL behavior for all the doping shown in Fig.~\ref{DOSvsmu}(a).
In addition, the FS changes from electron-like
(e-FS) to hole-like (h-FS) crossing the chemical potential $\mu_{LT}$~\cite{helena_prl}, as we can
observe in Fig.~\ref{AkwU35}. 

\begin{figure}[t]
\begin{center}
  \includegraphics[scale=0.33]
                 {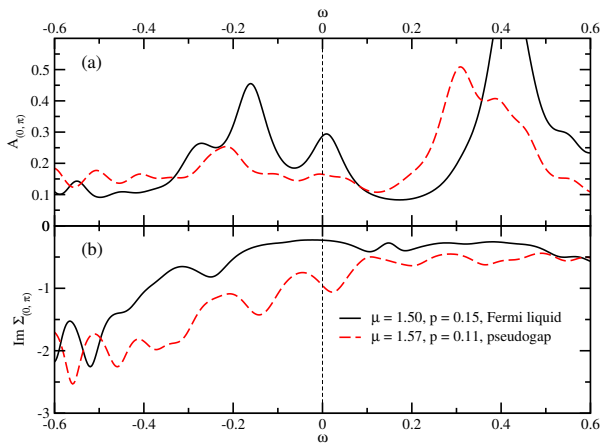}
\end{center}
\caption{(a) Spectral function and (b) imaginary part of the self-energy both at
  $\mathbf{k} = (0,\pi)$ as a function of frequency for $U=8.0$ and values of $\mu$ ($p$) very close to the FL-pseudogap transition. 
}
\label{GandS0P}
\end{figure}

The situation is very different for $U=8.0$ [see  Fig.~\ref{DOSvsmu}(b)]
as compared to $U=3.5$. For large $U$,
a discontinuous transition is seen between a FL
phase that exists for $\mu \le 1.50$ (corresponding to $p \ge 0.15$)
and a pseudogap phase, observed for $\mu \ge 1.57$ ($p \le 0.11$).
The transition is evident when we compare the spectral function and
the imaginary part of the self-energy, $\mbox{Im}\Sigma(\mathbf{k},\omega)$,
at $\mathbf{k}=(0,\pi)$ for $\mu=1.50$ and $\mu=1.57$, as depicted
in Fig.~\ref{GandS0P}. While for the former, the spectral function
presents a peak around $\omega = 0$ and $\mbox{Im}\Sigma_{(0,\pi)}
\propto \omega^2$ at low frequencies, characteristic
of a FL, for the latter no peak is seen in $\mbox{A}_{(0,\pi)}$ and
a non-FL feature is observed close to $\omega=0$ for $\mbox{Im}\Sigma_{(0,\pi)}$,
defining a pseudogap state. Most interestingly, the FL-pseudogap
transition is tied to a change in the FS topology: the pseudogap phase
has a hole-like FS, as we argued in Ref.~\cite{helena_prl},
while our results indicate that, in the strongly correlated regime at large
$U$, the FL phase has an
electron-like FS. 
We shall discuss this in more detail in section \ref{2B}.

As we mention in the introduction,
the coincidence between the doping $p^*$ up to which the
pseudogap is observed and the critical doping $p_{LT}$ of
the Lifshitz transition for the case of 
strong correlations has been observed in previous CDMFT
calculations for the Hubbard model~\cite{helena_prl,wu_2018}
and is in agreement with experimental
results~\cite{benhabib_2015,loret_2017,doiron_2017}.
What is interesting to learn from our new results is that
the local DOS at the Fermi level (and the nematic susceptibility, as
we will discuss next) at strong correlations is asymmetric
and changes discontinuously at $p_{LT}$ or $\mu_{LT}$, a
behavior that we ascribe to the FL-pseudogap phase transition,
not present for the case of weak correlations.

\subsubsection{Nematic susceptibility results}

We now turn to the $B_{1g}$ nematic susceptibility extracted from Raman response data.
Our interest is motivated in part by the 
experimental findings reported in Ref.~\cite{benhabib_2015, natphys_gallais}.
Here the authors focused on the Raman response for the cuprate
Bi$_2$Sr$_2$CaCuO$_{8+\delta}$. In the analysis of Ref.~\cite{natphys_gallais},
they identified the integral 
of Eq.~(\ref{hilbertw0}) 
$\chi^{\Lambda}(T)
=(1/\pi) \int_0^{\Lambda} d\nu \chi''(\nu)/\nu$,
where $T$ is the temperature, as a nematic susceptibility.
To be more
precise, 
they show results for the ratio between the integrals for
the $B_{1g}$ response obtained at a given $T$ and at $T=300$~K,
$\chi^{\Lambda}(T)/\chi^{\Lambda}(T=300\mbox{ K})$.
Around 100~K, which is above the critical temperature for all samples,
this ratio presents a non-monotonic behavior: for samples where a
pseudogap is not seen, the ratio increases as doping decreases, it
then reaches a maximum for the sample with $p \approx p^*$,
and it finally decreases as $p$ decreases even further, for samples
that present a pseudogap phase.

\begin{figure}[t]
\begin{center}
  \includegraphics[scale=0.33]
  {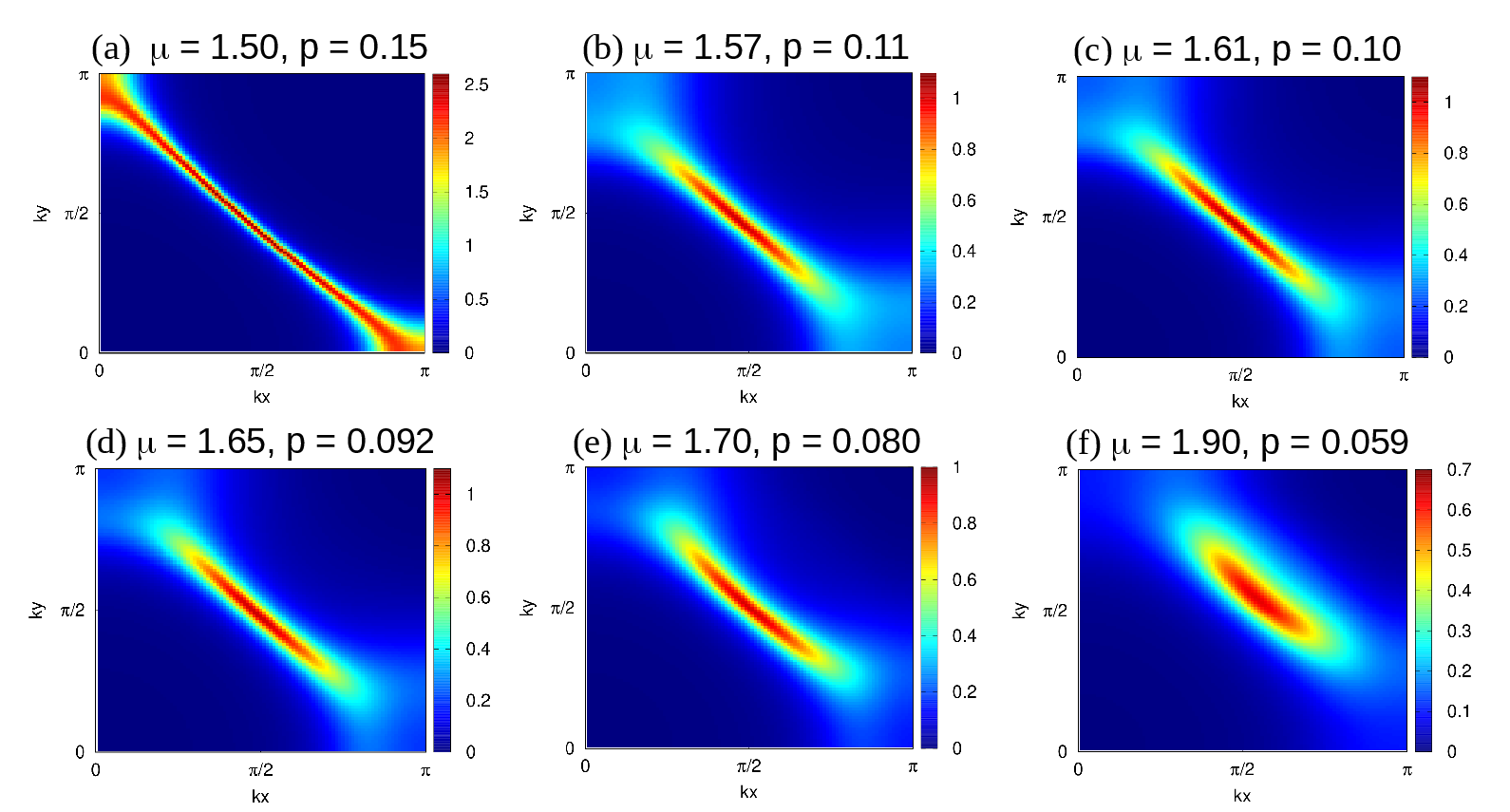}
\end{center}
\caption{
  $B_{1g}$ nematic susceptibility extracted from Raman response data as a function of the chemical
  potential $\mu$ for (a) $U=3.5$ and (b) $U=8.0$. The dashed red line in panel (a)
  denotes the Lifshitz transition. See the insets of Fig.~\ref{DOSvsmu} for the
  $n \times \mu$ plots corresponding to the results in this figure.}
\label{chivsmu}
\end{figure}

Our CDMFT results for the $B_{1g}$ nematic susceptibility,
$\chi_{B1g}(i\omega \rightarrow 0)$, are displayed in
Fig.~\ref{chivsmu}. The data are presented for the same parameters
as those used in Fig.~\ref{DOSvsmu}: in panel (a) we have results
in the weak coupling regime, $U=3.5$, and in panel~(b) those 
for strong coupling, $U=8.0$.
Remember that $\chi_{B1g}(i\omega \rightarrow 0)$ can be related
to the experimental accessible response, $\chi''(\nu)$, through
Eq.~(\ref{hilbertw0}). This means that the non-monotonic behavior
of $\chi_{B1g}(i\omega \rightarrow 0) \, \text{vs.} \, \mu$ observed in
Fig.~\ref{chivsmu}(b) is in general agreement to what is seen
experimentally in the cuprates and we described in the previous
paragraph.
Furthermore, the qualitative behavior followed by the nematic susceptibility
resembles that of the local DOS: for $U=3.5$, both
quantities are symmetrical with respect to $\mu_{LT}$, while for
$U=8.0$ we obtain a discontinuous and asymmetrical behavior.
This is the main result of our current work: we observe a change in
the nature of the Lifshitz transition, which becomes 
discontinuous when it is tied to the FL-pseudogap transition,
as seen in our local DOS and in the nematic susceptibility data for $U=8.0$. 

We note that our numerical results in the strong coupling regime -- namely, the non-monotonic behavior of the B$_{1g}$ nematic susceptibility as a function of doping, in particular the asymmetry around the Lifshitz transition point 
-- are consistent with measurements that suggest a connection between the pseudogap regime and nematic fluctuations. More specifically, the order parameter of the nematic phase in the square lattice has B$_{1g}$ symmetry and was investigated in Ref.~\cite{natphys_gallais}; the authors have observed enhanced nematic fluctuations near the pseudogap endpoint, as described above. Previously, scanning tunneling microscope experiments on Bi$_2$Sr$_2$CaCuO$_{8+\delta}$ samples have indicated the weakening of symmetry-breaking tendencies as the doping increases, thus suggesting a direct link between the Lifshitz transition and the disappearance of this symmetry-breaking close to the critical doping~\cite{stm_fujita_2014}.
In addition, divergent nematic susceptibility near the pseudogap critical point have been observed on (Bi,Pb)$_2$Sr$_2$CaCu$_2$O$_{8+\delta}$ samples through elastoresistance measurements~\cite{ishida_science_2020}.

The different behaviors we observe in the local DOS could be probed by spectroscopic
techniques such as STS (see, e.g., Ref.~\cite{piriou_2011}), provided materials with different correlation strengths are compared. In most cuprates (Hg-, Bi-based, and 
La$_{2-x}$Sr$_x$CuO$_4$), the Lifshitz transition occurs in a strongly correlated regime dominated by pseudogap physics and competing orders. By contrast, Tl$_2$Ba$_2$CuO$_{6+\delta}$ lies on the more weakly correlated side, and angle-resolved photoemission spectroscopy (ARPES) studies have established a doping-induced Lifshitz transition in this compound~\cite{Plate2005, Peets2007}. While scanning tunneling microscopy studies on Tl-2201 is notoriously difficult due to surface instabilities, Raman spectroscopy offers a bulk probe that has already been successfully applied to this material~\cite{Gasparov1998PRB, rmp_raman}, making it a promising route to test our
predictions in a regime closer to weak correlations.

\subsection{Fermi surface and Green's functions zeroes evolution
  in the pseudogap phase} \label{2B}

It is well established from ARPES spectra on various cuprates~\cite{damascelli_2003} that in going from the overdoped to
the underdoped side of the cuprate phase diagram a large FL-like FS breaks into pieces, known as as Fermi arcs. CDMFT studies~\cite{PhysRevLett.95.106402,BGK06} have proven that this phenomenon can be well accounted for as the correlation effect coming from doping a Mott insulator. Subsequent studies have pointed out that this phenomenon can be understood in terms of a novel kind of doping-driven Mott transition, which, differently from the usual one describing the transition between a FL metal and a Mott insulator, describes the transition from a FL metal into a pseudogap 
metal~\cite{Sordi2010_PRL, Sordi2011_PRB, helena_prl,Wu2018PseudogapTopology}.

\begin{figure}
\begin{center}
  \includegraphics[scale=0.33]
                  {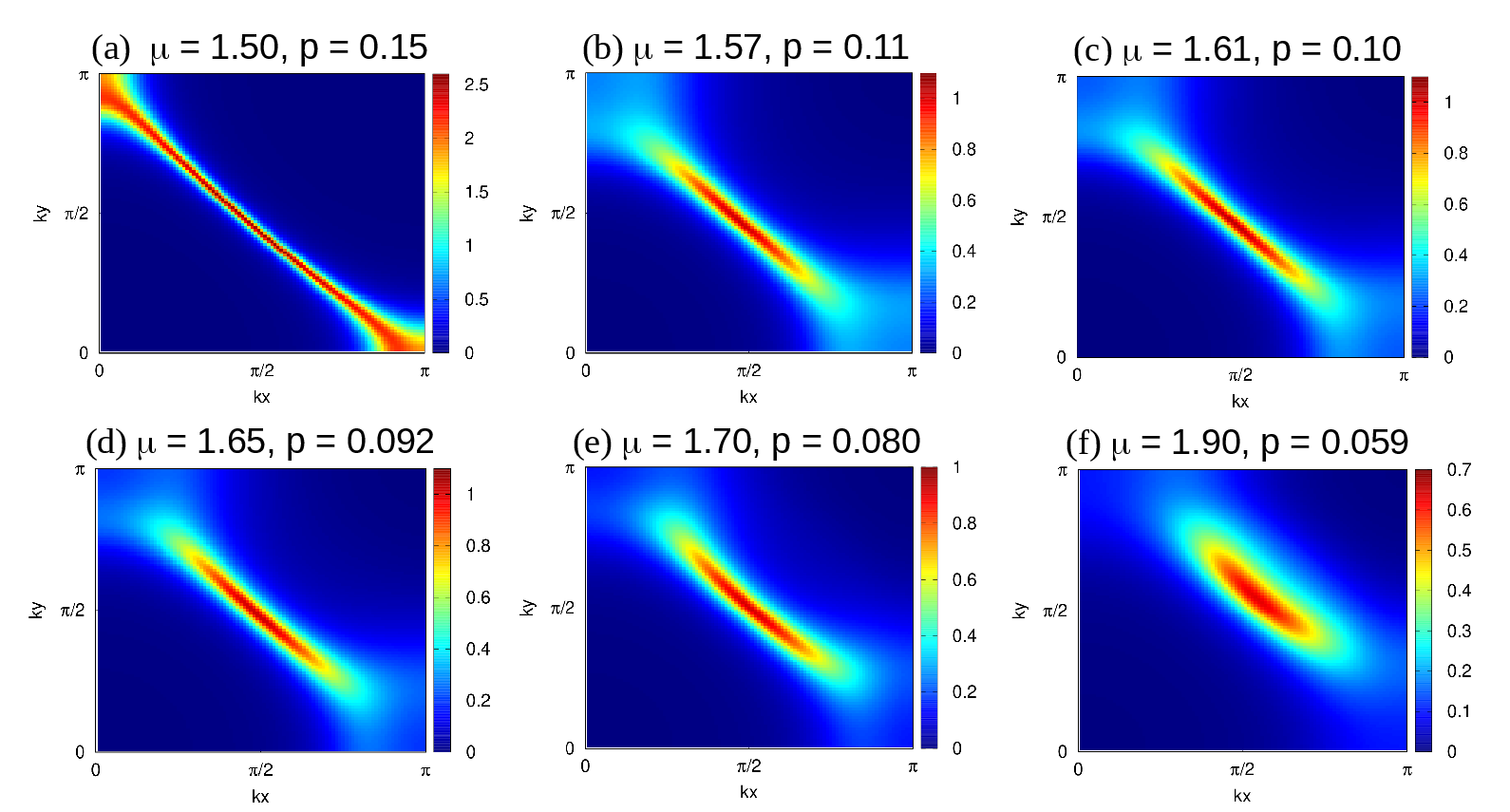}
\end{center}
\caption{Evolution of the spectral function $A(\mathbf{k}, iw \rightarrow 0$) for $U=8.0$ and different values of the chemical potential/doping, as indicated in each panel. In panel (a), the system in the FL phase; in the others, it is in the pseudogap phase.} 
\label{AKw1}
\end{figure}

In Fig.~\ref{AKw1} we display the CDMFT spectral function $A(\mathbf{k}, i\omega \rightarrow 0)$ from the overdoped to the underdoped side and we show that indeed this spectral intensity well portrays the physical picture of the FS breaking into arcs described in ARPES experiments~\cite{damascelli_2003}. In the following, we aim to discuss how, within our CDMFT approach, these spectral functions are related to the appearance of singularities in the Green's function known as ``zeroes''~\cite{PM}. 
We shall show that 
the key role in the change of FS topology is played by lines of
zeroes of the Green's function, where the properties of the system become non-FL.

Fermi liquids are characterized by the presence of poles in the Green's function,
which determine the FS lines in momentum space. In strong coupling systems however
zeroes of the Green's functions (divergences of the self-energy) can also appear.
For instance, the absence 
of Green's function poles inside a Mott insulating gap, replaced by lines of zeroes 
of the Green's function, is the typical hallmark of a Mott insulator. In the 
doped Mott insulator,
a coexistence of both poles and zeroes of the Green's function  
is at the origin of the pseudogap phase~\cite{PM,sakai2009evolution,sakai2010doped,FSslab2025}.
Note that in the doped Hubbard model phase diagram~\cite{helena_prl}, the pseudogap phase
  is observed for large values of $U$; for weak interactions, the system ground state is a FL.

\begin{figure}
\begin{center}
  \includegraphics[scale=0.38]
                  {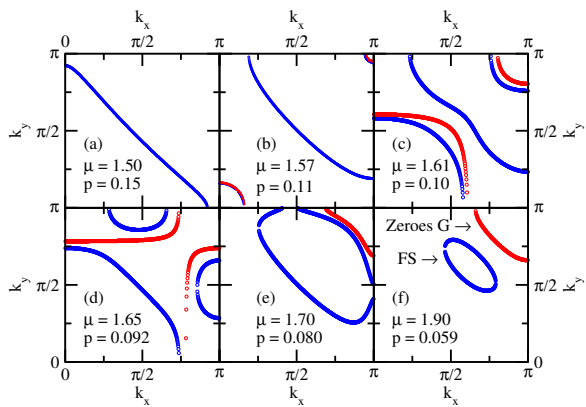}
\end{center}
\caption{Evolution of the FS (in blue) and zeroes of
  $G(\mathbf{k}, i\omega\rightarrow0)$ (in red) for $U=8.0$ and the same
  values of the chemical potential/doping as in Fig.~\ref{AKw1}.
  In panel (a), the system in the FL phase; in the others, it is in the pseudogap phase.}
\label{FSandzeros}
\end{figure}

To determine the poles and the zeroes lines of the Green's function, it is useful to define
the renormalized energy, $r(\mathbf{k}, i\omega)=\mbox{Re}[-1/G(\mathbf{k}, i\omega)]=
\mbox{Re}[\Sigma(\mathbf{k}, i\omega)]+\xi_{\mathbf{k}}$ (see
Appendix~\ref{renener} for plots of $r(\mathbf{k}, i\omega)$ for
a chosen value of $\mu$). The lattice Green's function $G(\mathbf{k}, i\omega)$ is determined 
with the cumulant periodization (detailed in Appendix~\ref{cumulant}).
The FS is given by the zeroes of $r(\mathbf{k}, i\omega)$ at the
Fermi level ($i\omega\rightarrow0$ limit), while the Green's function
zeroes correspond to the divergences of $r(\mathbf{k}, i\omega\rightarrow0)$.
Results for $U=8.0$ and different values of the chemical potential/doping
are shown in Fig.~\ref{FSandzeros} in the first quadrant of the BZ: the FS is represented by the blue curves, while the Green's function zeroes are in red. We observe a fast changing of such lines within the pseudogap phase.

In panel (a) ($\mu=1.50$, $p=0.15$), we show results in the more conventional FL phase, in which we observe a large, electron-like FS. No zero of the Green's function is present in this case.
At the emergence of the pseudogap phase $p=0.10-0.11$ [panels (b) and (c)], pockets of lines of zeroes accompanied by lines of poles of the Green's function appear around the $(0,0)$ and $(\pi,\pi)$ corner points, together with the large FS that one can observe close to diagonal of the quadrant (blue line). While mathematically this is the way the CDMFT implemented with the cumulant periodization describes the appearance of the lines of zeros, one must stress that these are not quantities physically observable within a spectroscopic probe. The effect of the lines of poles is to produce a quasiparticle (in principle visible), but the concomitant presence of the lines of zeroes reduces the weight of these poles to an infinitesimal unobservable value. However, the presence of the line of zeroes centered around $(0,0)$ does affect the spectral weight of the {\it physical} FS line close to the quadrant diagonal, producing a strong modulation in momentum space. The resulting spectral density is a strongly modulated FS, as we display in  the corresponding quadrants of Fig.~\ref{AKw1}.

By decreasing doping $p=0.092 - 0.080$ [panels (d) and (e)], the lines of zeroes plus lines of poles centered at the quadrant corners increase and merge,
producing other FS pockets, which are once again not physically meaningful, because they have very small spectral weight. The net effect is that the large FS realized at higher doping is replaced by a new FS, having stronger spectral weight close to so called nodal point ${\bf k}= (\pi/2,\pi/2)$ (see Fig.~\ref{AKw1}), as expected from ARPES. At small doping [$p=0.059$, panel (f)], the FS line closes to form a pocket centered around the nodal point, but with the upper side spectral weight strongly suppressed by the remaining zeroes arc centered around ${\bf k} = (\pi,\pi)$.

While a rather complicate interplay between lines of zeroes and poles of the Green's function takes place within our formalism, the physical picture is very simple: the large FS from high doping becomes the well-known Fermi arc at small doping, as portrayed in Fig.~\ref{AKw1}, 
as soon as the lines of zeroes appear at the Fermi level.

As the system approaches the insulating phase
($p=0$), the FS disappears and only the line of zeroes survives. 
Interestingly, a similar evolution of the FS and the Green's
functions zeroes has been reported previously for the normal component of the 
superconducting
state within CDMFT calculations~\cite{civelli_2009}.

Our results in Fig.~\ref{AKw1} and \ref{FSandzeros}, which are based on the microscopic 
resolution of the Hubbard Model within CDMFT, are in qualitative agreement
with other phenomenological
proposals~\cite{PhysRevB.73.174501,storey_2016,storey_2017,PhysRevB.107.L201121}. 
To be more specific, a reconstruction from a large hole-like FS into
a small elliptical pocket as the doping decreases in the pseudogap
phase was proposed based on the Yang, Rice, and Zhang model by
considering the folding of the BZ on the zone diagonals, due to
antiferromagnetic order. In Ref.~\cite{storey_2016}, this
reconstruction is shown to take place in the same doping range
as that in which the Hall number $n_H$ goes from $1+p$ to $p$
as the doping decreases. The dependence of $n_H$ with $p$ within
the cited theoretical treatments agrees well with experimental
results~\cite{badoux_2016,collignon_2017}. The experimental
$n_H \, \text{vs.} \, p$ data
are also well reproduced if the carrier density is calculated within
the same models by assuming a mean free path proportional to the
circumference of the hole-like FS pocket. In this case, the mean
free path is shown to drop abruptly when the pseudogap
emerges~\cite{storey_2017}. As we discussed in Sec.~\ref{DOSRaman},
the DOS at the Fermi level calculated by us within CDMFT [see
Fig.~\ref{DOSvsmu}(b)] also drops abruptly when the system enters the
pseudogap phase.

\section{Conclusions} \label{conclusions}

In the current paper, we have studied the Hubbard model in two dimensions using cellular dynamical mean field theory. We have looked into the behavior of the density of states (DOS) at the Fermi level and the nematic susceptibility extracted from Raman response data across the interaction versus doping phase diagram. The latter has been calculated in the B$_{1g}$ channel, the one that signals the pseudogap phase, and can be identified as a measure of nematic fluctuations. 
In the weakly correlated regime, where the system is in a Fermi liquid phase, both quantities are approximately symmetrical around the critical doping $p_{LT}$ where the topology of the Fermi surface changes (Lifshitz transition). On the other hand, in the strongly correlated case, relevant for cuprates, we observe a discontinuous and asymmetrical behavior when the system goes from the Fermi liquid phase to the pseudogap one as doping decreases. Thus, our results suggest a change in the nature of the Lifshitz transition when comparing the two electronic interaction regimes, which we associate to the fact that at strong correlations it is tied to the transition between the Fermi liquid and pseudogap phases. Within the latter, we also analyze how the Fermi surface evolves from a large hole-like one
close to $p_{LT}$ into a small elliptical pocket at small doping.

The asymmetrical behavior in the DOS and in the B$_{1g}$ nematic susceptibility for strong interactions, along with the change in Fermi surface topology upon entering the pseudogap phase, are consistent with current experiments on cuprates~\cite{natphys_gallais}, and should motivate further studies on compounds with varying correlation strength undergoing a Lifshitz transition, such as Tl$_2$Ba$_2$CuO$_{6+\delta}$ on the more weakly correlated side.

We acknowledge financial support from FAPEMIG, CNPq, and CAPES. M.C. acknowledges
support from the ANR grant NEPTUN no. ANR-19-CE30-0019-04.

\bibliographystyle{apsrev4-1}
\bibliography{references}

\appendix

\section{Periodization of cluster observables} \label{cumulant}

In cellular dynamical mean-field theory, a $2\times 2$ plaquette provides frequency-dependent cluster quantities 
$\mathcal{O}_{\mu\nu}(i\omega_n)$, with $\mu,\nu=1,\dots,N_c$ labeling intra-cluster sites. 
Typical examples include the cluster Green's function $G_{\mu\nu}$, which explicitly enters the CDMFT self-consistency condition.  
To recover translationally invariant lattice observables from these cluster objects, one usually resorts to a periodization procedure. The idea is to regard the cluster matrix elements as Fourier coefficients and reconstruct the momentum-resolved lattice quantity through a truncated Fourier series. For a cluster of size $N_c$, one writes
\begin{equation}
 \mathcal{O}^L(\mathbf{k},i\omega_n) = \frac{1}{N_c}\sum_{\mu,\nu=1}^{N_c}
 \mathcal{O}^c_{\mu\nu}(i\omega_n)\, e^{i\mathbf{k}\cdot(\mathbf{r}_\mu-\mathbf{r}_\nu)},
\end{equation}
where superscripts $L$ and $c$ denote lattice and cluster, and $\mathbf{r}_\mu$ are the intra-cluster positions.  

Several choices for the cluster observable have been explored. One possibility is to periodize the cluster self-energy~\cite{TeseMarcello2}, or the Green's function itself~\cite{GK1,SSK08}. Because of the truncation in the Fourier expansion, the best suitable quantities are the ones that are more local, i.e. better described 
within the cluster size.  
Such a local quantity is the cluster cumulant,
\begin{equation}
 \hat{M}^c = \left[(i\omega_n + \mu)\hat{I} - \hat{\Sigma}^c\right]^{-1},
\end{equation}
whose Fourier components decay faster and therefore yield a more reliable reconstruction~\cite{PM,PM2}.  
Comparative analyses confirm that the cumulant scheme captures most effectively the physics near the Mott transition and inside the insulating regime~\cite{SIZE}.  

Once the lattice cumulant $M^L(\mathbf{k},i\omega_n)$ is obtained, the lattice Green's function and self-energy follow directly as
\begin{equation}
 G^L(\mathbf{k},i\omega_n) = \left[M^L(\mathbf{k},i\omega_n)^{-1}-\xi_{\mathbf{k}}- \mu  
 )\right]^{-1},
\end{equation}
with $\xi_{\mathbf{k}}$ given in Eq.~\ref{dispersion},
and
\begin{equation}
 \Sigma^L(\mathbf{k},i\omega_n) = i\omega_n + \mu - M^L(\mathbf{k},i\omega_n)^{-1}.
\end{equation} 
In the main text, we drop the superscript $L$ for simplicity.

\section{Raman $\gamma(\mathbf{k})$ factor}  \label{app_raman}

Let us specify the Raman scattering amplitude $\gamma(\mathbf{k})$
that appears in Eq.~(\ref{chiiom}). It is given by~\cite{rmp_raman}
\begin{equation}
\gamma(\mathbf{k})=\sum_{\alpha,\beta} \gamma_{\alpha,\beta}(\mathbf{k})
e^{\alpha}_i e^{\beta}_j,
\end{equation}
where $e^{\alpha}_i$ denote the components of the vector defining the light
polarization direction. Considering the bubble contribution~\cite{rmp_raman}, we have
\begin{equation}
  \gamma_{\alpha,\beta}(\mathbf{k})=\frac{1}{\hbar^2}\frac{\partial^2 \xi_{\mathbf{k}}}
{\partial k_{\alpha} \partial k_{\beta}},
\end{equation}
where $\xi_{\mathbf{k}}$ is the electronic dispersion, defined in Eq. (\ref{dispersion})
in the present case.

For the $B_{1g}$ symmetry, incident and scattered light polarizations are
aligned along~\cite{prb_raman} \^x+\^y and \^x-\^y and
\begin{equation}
\gamma_{B1g}(\mathbf{k})=\frac{\partial^2 \xi_{\mathbf{k}}}
{\partial k_x^2} - \frac{\partial^2 \xi_{\mathbf{k}}}{\partial k_y^2}.
\end{equation}

In the case of the 2D Hubbard model we consider [whose dispersion is given in
Eq.~(\ref{dispersion})], we obtain Eq.~(\ref{ramanB1g}).

\begin{figure}
\begin{center}
  \includegraphics[scale=0.33]
                  {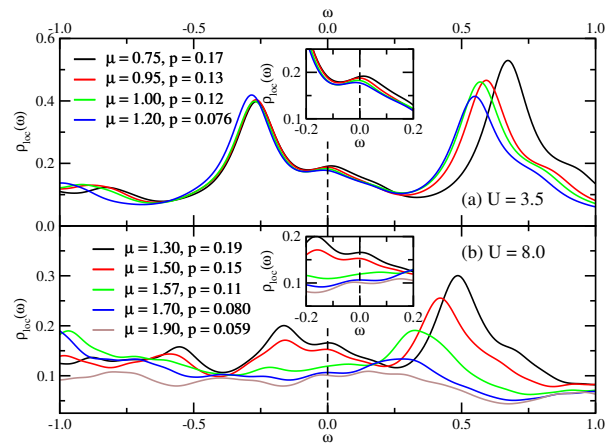}
\end{center}
\caption{Local DOS as a function of frequency $\omega$ for different values of $\mu$, $p$ for (a) $U=3.5$ and (b) $U=8.0$. The insets show a zoom in of the results in the main panels. These results were obtained on the real axis at $\omega + i \eta$, with a broadening parameter $\eta=0.1$, which justifies a (small) difference between the values observed in this figure at $\omega \approx 0$ and those of Fig.~\ref{DOSvsmu}, calculated on the imaginary axis, at the first Matsubara frequency.}
\label{DOSvsom}
\end{figure}

\section{Density of states as a function of frequency} \label{DOScomparison}

In Fig.~\ref{DOSvsom} we present the local DOS as a function of frequency for (a) $U=3.5$ and (b) $U=8.0$ for various $\mu$, $p$ values, from the overdoped to the underdoped sides. As pointed out in the main text, for $U=3.5$ the system is in the FL phase, while for $U=8.0$ it goes from the FL regime at small $\mu$ (black and red curves) to the pseudogap phase at $\mu \ge 1.57$, $p \le 0.11$ (green, blue, and brown curves). We observe a small peak around $\omega \approx 0$ at the FL phase, both for $U=3.5$ and $U=8.0$ at large doping, which is suppressed once the electronic interaction is large ($U=8.0$) and the system enters the pseudogap (small doping). 

Together with the appearance of the pseudogap, the system crosses a VHS at intermediate doping $p_{LT}$. In simple tight binding DOS, this should manifest as a weak (logarithmic) divergence crossing the Fermi level. In the CDMFT DOS, however, finite quasiparticle lifetime intrinsic in the correlated regime [Im$\Sigma(\omega)\neq 0$], even close to the Fermi level, washes away such divergence. It becomes therefore difficult to observe the VHS in physical scanning tunneling spectra, for instance. However the sign of the slope of the DOS close to $\omega=0$ does change sign while crossing the VHS (see the insets of Fig.~\ref{DOSvsom}).
This change of behavior around the Fermi energy is captured when plotting the local DOS at $i\omega \rightarrow 0$ as a function of $\mu$, as we did in Fig.~\ref{DOSvsmu} of the main text.

\section{Renormalized energy} \label{renener}

\begin{figure}
\begin{center}
  \includegraphics[scale=0.33]
                  {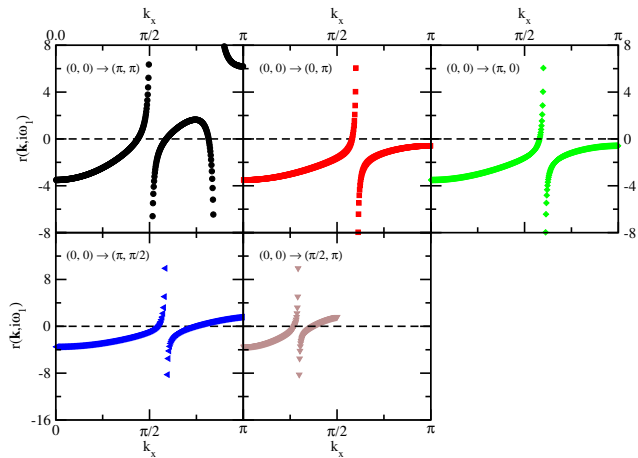}
\end{center}
\caption{Renormalized energy at the smallest Matsubara frequency as a
  function of $k_x$ for $\mu = 1.61$ and
  different directions in $\mathbf{k}$ space, as indicated in each panel.
  The FS and lines of zeroes corresponding to this value of the chemical
  potential are displayed in panel (c) of Fig.~\ref{FSandzeros}.}
\label{fig_renener}
\end{figure}

To exemplify the behavior of the renormalized energy defined
in the main text, it is presented in Fig.~\ref{fig_renener} 
at the first Matsubara frequency ($\omega_1 = 0.0314$)
and different directions in $\mathbf{k}$ space for the system
with $\mu = 1.61$.
As previously explained, it is from the zeroes and divergences
observed in $r(\mathbf{k}, i\omega \rightarrow 0)$ that the FS
and the zeroes of the Green's functions are determined.

\end{document}